\documentclass{emulateapj}  

\usepackage{epsf}
\usepackage{amsmath}
\usepackage{amssymb}
\usepackage{graphics}
\usepackage{graphicx}

\submitted{Accepted for publication in ApJ}

\def\MgII{Mg\,{\sc ii}}

\begin{document}

\title{Cosmic dust in \MgII\ absorbers}

\author{Brice M\'enard\altaffilmark{1,2,3}
 \& Masataka Fukugita\altaffilmark{2,4,5}}

\altaffiltext{1}{Department of Physics \& Astronomy, Johns Hopkins University,
Baltimore, MD 21218 U.S.A.}
\altaffiltext{2}{Institute for the Physics and Mathematics of the Universe,
University of Tokyo, Kashiwa 277-8583, Japan}
\altaffiltext{3}{Alfred P. Sloan fellow}
\altaffiltext{4}{Institute for Advanced Study, Princeton, NJ 08540 U.S.A}
\altaffiltext{5}{Institute for Cosmic Ray Research, University of Tokyo, Kashiwa 2778582, Japan}

\begin{abstract}

\MgII\ absorbers induce reddening on background quasars.
We measure this effect and infer the
cosmic density of dust residing in these systems to be
$\Omega\approx 2\times 10^{-6}$, in units of the critical density of
the universe, which is comparable to the amount of dust found in
galactic disks or about half the amount inferred to
exist outside galaxies.  We also estimate the neutral hydrogen
abundance in \MgII\ clouds to be $\Omega\approx 1.5\times 10^{-4}$,
which is approximately 5\% of hydrogen in stars in galaxies. This
implies a dust-to-gas mass ratio for \MgII\ clouds of about $1/100$,
which is similar to the value for normal galaxies. This would
support the hypothesis of the outflow origin of \MgII\ clouds, which are
intrinsically devoid of stars and hence have no  sources of dust.
Considerations of the dust abundance imply that the presence of
\MgII\ absorbers around
galaxies lasts effectively for a few Gyr. High redshift absorbers
allow us to measure the rest-frame extinction curve to $900\,{\rm
\AA}$, at which the absorption by the Lyman edge dominates over
scattering by dust in the extinction opacity.

\end{abstract}

\keywords{dust -- extinction -- galaxies: halos -- quasars: absorption line}

\section{Introduction} 
\label{sec:introduction}

A significant amount of gas processed by stars may be ejected into
interstellar space via stellar winds and supernova explosions.  About
30\% of the metals in enriched gas would condense to form dust
grains \citep{2001ApJ...548..296W}.  The total amount of dust in the
Universe that is produced in stellar evolution in the entire cosmic
time was estimated from integrated star formation rate by
\citet{fukugita2011} to be $\Omega_{\rm dust} \simeq 1\times
10^{-5}$, in units of the present-day critical mass density.
This value is a few times
higher than the amount of dust observed in galactic disks
\citep{2004ApJ...616..643F,2007MNRAS.379.1022D}, which motivated us to
explore the fate of the remaining amount.

Evidence for the existence of dust beyond galactic disks has
observationally been suggested by a few authors. From the study of a
low-redshift foreground/background galaxy superposition
\citet{2009AJ....137.3000H} detected dust extinction up to about five
times the optical extent of spiral galaxies. Using deep
\emph{Herschel} observations of M82, \citet{2010A&A...518L..66R}
showed that emission from cold dust can be traced up to 20 kpc from
the centre of the galaxy. Recently, \citet[hereafter MSFR]{2010MNRAS.405.1025M} measured the cross-correlation between the colors of
distant quasars and foreground galaxies as a function 
of the angular separation 
to galaxies, and
concluded an excess reddening signal on scales ranging from 20 kpc to a
few Mpc, implying the existence of a appreciable amount of dust in 
intergalactic space. Using this observational result,
\citet{fukugita2011} showed that the summed contributions of
dust in and outside galaxies appear to be in agreement with the total
amount of dust ought to be produced in the Universe. This implies that dust
destruction does not play a major role in the global dust distribution
and that most of the intergalactic dust survives over the cosmic time.

In this paper we pursue another line of observations and show
that a significant amount of dust resides in galactic halos and possibly
beyond: we use \MgII\ absorbers to find dust contained therein. 
\MgII\ is the most commonly detected absorption line from
cool gas ($T\sim 10^4$ K) at $z<2$ in the optical spectra of
distant sources. Strong \MgII\ absorbers, conventionally defined
with a rest equivalent width $W_0^{\lambda 2796}>0.3\,{\rm \AA}$, are
associated with a range of galaxies with $L\gtrsim 0.1 \,L^\star$
\citep{1991A&A...243..344B,Steidel-Sargent,SDP94,Nestor07}
and are found at impact parameters ranging 
up to 100 kpc \citep{SDP94,Steidel+97,Zibetti+07}.
While the physical
mechanisms for the origin of these gas clouds are yet to be understood,
most \MgII\ absorbers seem to reside in galactic haloes.

The use of absorbers offers an attractive property: the knowledge of
the absorber incidence $d N/dz$ allows us to infer the cosmic mass
density of dust contained in these systems without further assumptions
on their spatial distribution.  \MgII\ absorbers can hence be used to
obtain a robust lower limit on the amount of baryons and dust residing
outside galactic disks up to the halo radius, 
over the redshift range $0.5\lesssim z\lesssim 2$.
In general, it is widely believed that \MgII\
absorbers do not host star forming regions and hence no source of dust grains.
The presence of dust in these clouds would serve as an indicator to distinguish whether \MgII\ absorbers predominantly are aggregates of pristine gas or are of
secondary products from activities of nearby galaxies.
Characterizing and understanding its distribution is therefore an important
task.
The use of absorbers also allows us to
infer the neutral hydrogen abundance in \MgII\ clouds.  These pieces of
information concerning the ingredients of the clouds would lead us to infer
properties of \MgII\ absorbers, and then the nature and the formation
of these systems.  At the same time this study enables us to explore
the feature of the extinction curve to the short wavelength due to the
high redshift nature of clouds, even to
beyond the Lyman limit, using optical and UV data currently available.

Our basic data are taken from the SDSS (York et al. 2001) 
but are supplemented with UV photometry of 
the Galaxy Evolution Explorer (GALEX; Martin et al. 2005). 
We use $H_0=70$ km s$^{-1}$Mpc$^{-1}$ and $\Omega_M=0.3$ in a flat Universe.

\section{Expected optical and UV absorption by \MgII\ absorbers}  
\label{sec:theory}

Metal-enriched gas causes extinction of light from UV to near infrared
passing through intergalactic matter due to absorption and scattering
by dust grains.  In the far UV region we also anticipate that
ionisation and excitation of hydrogen atoms lead to an attenuation of
light. This is usually not addressed
in the context of dust extinction 
observed in optical wavelengths.
In a similar manner excitation of heavy
elements may also contribute to the extinction opacity.

We write the optical depth for extinction as
\begin{equation}
  \tau(\lambda) = 
  N_{\rm HI}
\left[
  \sigma_{\rm H}(\lambda) +
  \sum_i\sigma_{{\rm d},i}(\lambda)\,\delta_i
    + (M/H)\sigma_M (\lambda)
  \right]
\label{eq:tau}
\end{equation}
where $\sigma_{\rm H}(\lambda)$ is the hydrogen cross section,
$\sigma_{{\rm d},i}$ is the dust extinction cross-section per hydrogen
atom for a given population of dust grains denoted by $i$, 
$\delta_i=(A_V/N_{\rm H}) / (A_V/N_{\rm H})_i$ accounts for variations
in the dust-to-gas ratio, 
$\sigma_M$ is the metal excitation cross section\footnote{We neglect 
the effect of line saturation at this stage.}
and $M/H$ is the metallicity in the ratio of numbers of atoms.

\begin{figure}
   \centering
   \includegraphics[width=\hsize]{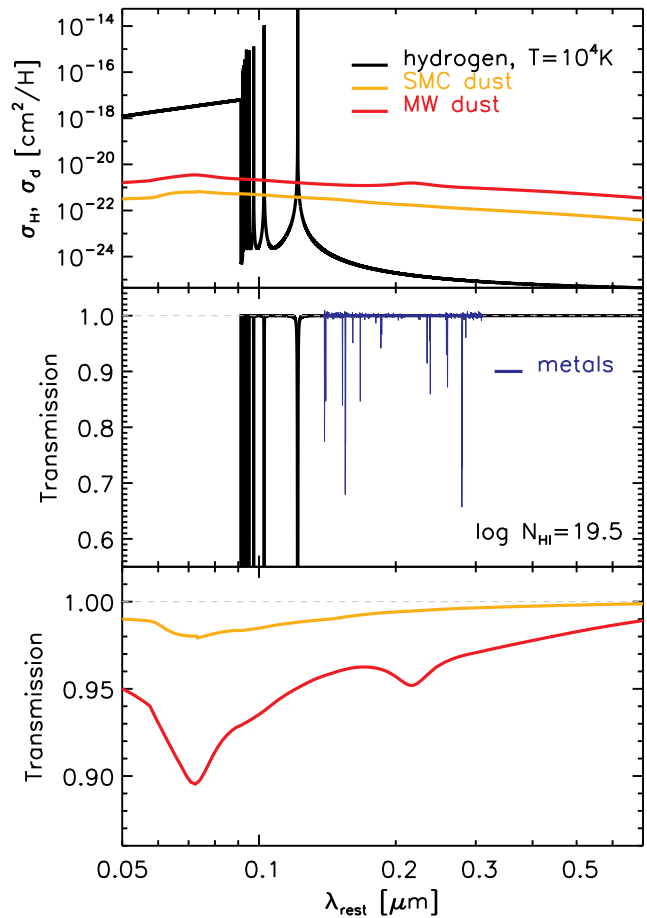}
   \caption{ \emph{Top:} 
    absorption cross-section of hydrogen atoms 
     (in black) and extinction cross section of
     dust grains of SMC type  (in orange) and Milky
     way type (in red), estimated by \citet{2001ApJ...548..296W}.
      \emph{Middle:} 
      the black curve shows the transmission through neutral hydrogen for 
     $N_{\rm HI}=10^{19.5}\,{\rm cm^{-2}}$ and $T=10^4$K.
     The blue curve shows a composite metal absorption spectrum for \MgII\ absorbers.  \emph{Bottom:}
      transmission through Milky Way type and SMC
     type gas with the corresponding dust-to-gas ratios.
     }
   \label{fig:cross-section}
\end{figure}

Figure~\ref{fig:cross-section} indicates the cross-sections as a
function of wavelength from the UV to the optical regime in (a), and
the representative transmission for a cloud with hydrogen column
density $N_H=10^{19.5}\,{\rm cm^2}$
in (b) and (c).  Dust models for the Milky Way and
the SMC  are adopted from \citet{2001ApJ...548..296W}\footnote{ Data
  for these models are taken from
  http://www.astro.princeton.edu/$\sim$draine/dust/dustmix.html }.
These models, consisting of mixture of carbonaceous grains and
astronomical silicate grains, reproduce the observed extinction curves
from the ultraviolet to the near infrared in the Milky Way and
Magellanic Clouds. They are normalized such that $(A_V/N_{\rm H})_{\rm
  MW}=5.3\times10^{-22}\,{\rm cm^2}$ and $(A_V/N_{\rm H})_{\rm
  SMC}=6.2\times10^{-23}\,{\rm cm^2}$, respectively, where
$A_V$ stands for extinction in the $V$ band.  
The hydrogen cross-section is shown for gas at a temperature 
$T=10^4$K, but other choices do not modify the  
resonant feature qualitatively\footnote{We thank 
Jens Chluba for providing us with estimates of the hydrogen cross-section.}.
There are humps at $\lambda\simeq 0.217\,\mu$m and $0.072\,\mu$m in
the dust extinction curve caused by carbonaceous grains, though they
are squeezed to a barely recognizable level in the top panel of
the figure due to the broad logarithmic scale.

The transmission $T=1-e^{-\tau}$ shown in the lower panels of
Figure~\ref{fig:cross-section} assumes the neutral hydrogen column density
$N_{\rm HI}=10^{19.5}\,{\rm cm^{-2}}$, which corresponds roughly to the median
column density of \MgII\ absorbers with $W_0=1\,{\rm \AA}$
\citep{2009MNRAS.393..808M}.  Equation~\ref{eq:tau} shows that we can
obtain, in principle, information on $\delta_i$ and $N_{\rm
  HI}$ from the brightness of a source located behind such a cloud.  The
figure shows that atomic excitation not only at the Lyman edge but
also in the Lyman region is a source of optical depth.  A cloud of
this hydrogen column density is completely opaque at wavelengths
shorter than the Lyman limit, $\lambda\simeq 911.75\,$\AA. At longer
wavelengths the optical depth is $\tau\lesssim 0.1$, which is dominated by
dust extinction. A drop at 2175\AA~is seen in the transmission
through Milky Way type dust.

\begin{figure} 
   \centering
   \includegraphics[width=1.0\hsize]{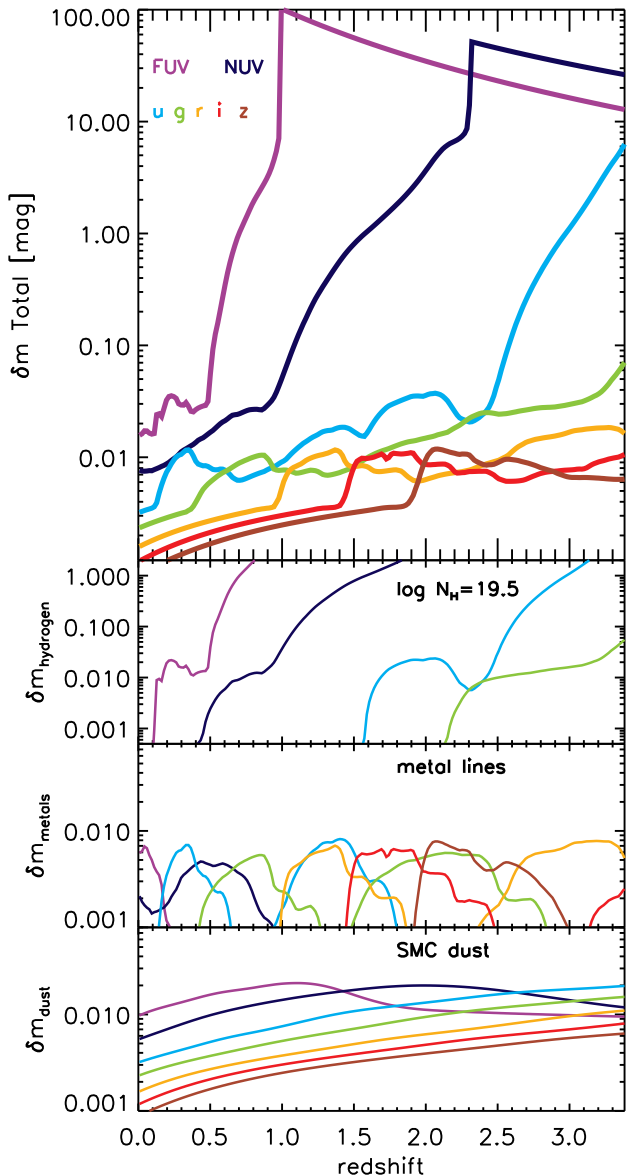}
	\vspace{-0.6cm}
   \caption{Flux attenuation induced by intervening dusty gas with
      $N_{\rm HI}=10^{19.5}\,{\rm cm^{-2}}$ as a function of
     redshift, using the SMC type dust model from
     \citet{2001ApJ...548..296W}. Breakdown into hydrogen, metal excitation
  and dust is shown in the three lower panels. Seven colour pass bands 
are shown
with different colours indicated in the legend.}
   \label{fig:mag_shift}
\end{figure}

We show an example of extinction for observations conducted with broad
band filters in Figure~\ref{fig:mag_shift}.
The upper panel shows the attenuation of the flux
induced by SMC-type dusty gas of hydrogen column density $N_{\rm
  HI}=10^{19.5}$cm$^{-2}$, as a function of redshift for the SDSS
($u,g,r,i,z$) and GALEX (NUV, FUV) passbands.  
Individual contributions are shown in the lower panels:
 (b) hydrogen, (c) metal lines, and (d) dust grains.  
 The seven passbands are denoted by different colours, as
specified in the legend of the figure.  
Lyman-edge absorption dominates in the FUV, NUV and $u$ bands
above redshifts 0.5, 1 and 2.5, respectively, as seen when one compares (a) 
and (b).

We include atomic excitations from metal lines using a composite 
absorption spectrum for \MgII\ clouds.
\footnote{We thank Guangtun Zhu for his help in creating the metal-line composite spectrum.}
The total contribution to the absorption opacity, which is dominated by
\MgII, {Fe\,{\sc ii}} and {C\,{\sc iv}} lines, 
can be comparable to that of Lyman-$\alpha$.  The 0.01 mag hump seen in the $u$ band  at $z=1.7-2.3$ 
is due to Lyman line absorption and the feature at $z\sim 0.3$ is due to
absorption by metals lines.
The Lyman absorption is also seen
in the FUV and NUV passbands, although the feature is somewhat weakened in NUV
due to its larger width of the bandpass
($\Delta\lambda\sim2000\,{\rm \AA}$). 
At $z>2.0$ we see that Lyman edge absorption becomes stronger 
in the NUV than the FUV band. This causes a blueing effect in these bands. Similarly, this is expected for  the NUV$-u$ color at $z\gtrsim 3.5$.

Absorption due to metal excitation is also visible
in the curves for $g,r,i,z$ passbands. We see that metal excitation
increases the extinction opacity from dust alone by as much as 5 times in the
relevant wavelength range in all optical passbands.
With the inclusion of metal excitation opacity the extinction
curve is not always monotonic, and the colour indices receive 
occasionally blueing for some wavelength ranges rather than
reddening, with an amplitude of the order 0.01 magnitude.

Choosing Milky Way type dust in our demonstration would
increase the contribution from dust opacity by 
the relative $(A_V/N_{\rm H})$ ratio, i.e. by roughly a factor 8.
Here, our choice of the SMC type dust is motivated from the fact that
it fits the extinction spectrum of \MgII\ absorbers better,
especially at short wavelengths 
$\lambda<2000\,$\AA, that is important to our consideration, 
and for the absence of the $\lambda$2175 
feature \citep{2006MNRAS.367..945Y}.
When interpreting observations we leave, however, the dust-to-gas ratio as a free parameter, rather than fix it to the actual SMC value. 
This ratio is constrained from both shape of the
extinction curve, for extinction is caused by both Lyman edge
absorption and dust scattering, and amount of dust and HI gas
associated with \MgII\ absorbers.

Previous analyses based on optical data have constrained the amount of
dust in intervening \MgII\ absorbers by statistically measuring 
color changes they induce on background sources
\citep{2006MNRAS.367..945Y,2008MNRAS.385.1053M,2011MNRAS.416.1871B}.
Here we extend to the UV, which enables us to obtain
information also on the amount of neutral hydrogen associated with \MgII\ absorbers.

\section{Analysis for \MgII\ absorbers}  \label{sec:analysis}

\subsection{Data}

The SDSS (York et al. 2001) provides us with the basic data set for the present analysis.
\citet{Nestor+05} constructed a catalogue of \MgII\ absorbers
using its EDR data base \citep{2002AJ....123..485S}, 
and \citet{2011AJ....141..137Q} extended it to the DR4 dataset \citep{2006ApJS..162...38A}.  \citet{2011AJ....141..137Q} analyzed about 45,000 quasar spectra and  identified
about 17,000 \MgII\ absorbers. \citet{Nestor+05} estimated
that the redshift path covered by the survey
drops to about 50\% for absorbers with the rest-system equivalent width $W_0<0.8$\AA. While completeness is not an important issue in our analysis, we restrict our study to the sample of absorbers with $W_0>0.8$\AA.
We do not include absorption lines located close to the edges of the spectra and restrict the \MgII\ absorber redshift distribution to $0.4<z<2.1$.
It is known from studies of absorber-galaxy pairs (e.g., Steidel
et al. 1994) that MgII absorber is unlikely to be a part of galactic
disks, but are placed substantially away from galaxies.
 
About 70\% of the SDSS quasars are observed by
GALEX  (Martin et al. 2005). 
Budavari et al. (2009) cross-matched the 
SDSS-DR7 catalog with the GALEX-GR4/5 taking a
matching radius of 4$''$. 
Using this cross-identification, we can find 11,929 \MgII\ absorption systems
satisfying our rest equivalent width and redshift selections and
 for which UV observations of the background quasars are available. 
This constitutes the prime sample of our analysis.

\subsection{Dust extinction}

\begin{figure}
   \centering
   \includegraphics[width=1.\hsize]{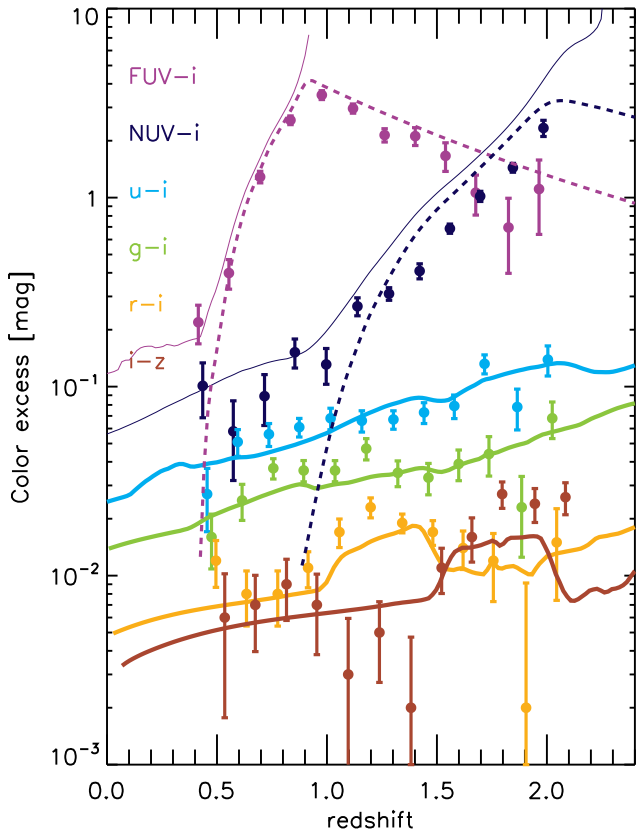}
   \caption{Reddening induced by the presence of \MgII\ absorbers 
with $W_0>0.8\,$\AA, as a
function of redshift for  FUV$-i$, NUV$-i$, $u-i$, $g-i$, $r-i$ and $i-z$ colours
in our quasar sample. The solid lines are reddening expected for
clouds with SMC type dust having
$N_{\rm HI}=10^{19.8}$ cm$^{-2}$
but scaled to the dust-to-hydrogen ratio of 1/100 consistent with our determinations given later.
Dashed curves are expectation for $N_{\rm HI}=10^{18}$ cm$^{-2}$ clouds.
}
   \label{fig:color_redshift}
\end{figure}

\begin{figure} 
   \centering
  \includegraphics[height=\hsize]{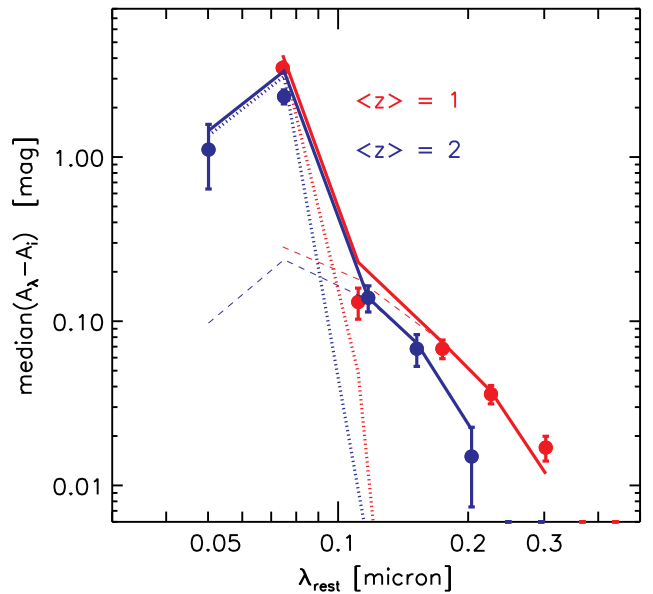}
  \caption{Median reddening curve $A_\lambda-A_{i}$ for \MgII\
    absorbers at $z\simeq 1$ and 2 (solid circles).
   Reddening is evaluated at the median wavelengths of the 5 filters
shorter than the $r$ band. 
   The dashed curve is the dust extinction curve of 
\citet{2001ApJ...548..296W} for SMC, and dotted is absorption by hydrogen. The two  curves
are added to give the total extinction curves shown by thick solid curves.
The curves assume $N_{\rm HI}=10^{19.8}$cm$^{-2}$ with the dust-to-gas
ratio 1/100. The hydrogen absorption is evaluated for 
$N_{\rm HI}=10^{18}$cm$^{-2}$.
}
   \label{fig:reddening_shape}
\end{figure}

To detect dust in \MgII\ absorbers we measure reddening in broad band
brightness correlated with the presence of absorption systems in
quasar spectra.  For each quasar with a detected \MgII\ absorber we
select four nearest neighbour quasars in redshift--$g$-band magnitude
space and take them to be reference quasars.  
The high density of
points in this space allows us to find reference quasars within a
brightness difference typically smaller than 0.1 mag in the $g$-band.
We then compute a median colour excess and estimate its error by bootstrapping the absorber sample.

Figure \ref{fig:color_redshift} shows the reddening due to all
\MgII\ absorbers with $W_0>0.8\,$\AA\ as a function of redshift.  For
six colours, FUV$-i$, NUV$-i$, $u-i$, $g-i$, $r-i$ and $i-z$, the solid curves
show the expected reddening including hydrogen absorption for log
$N_{\rm HI}=19.8$ with SMC-type dust but with 
a dust abundance lowered by a factor 6, corresponding to a
dust-to-gas mass ratio of about 1/108 
Our reddening estimation also includes absorption from metal lines.  The overall agreement is good:
the deviations from the theory curve are of the order of a few percent except
for some specific case, which we discuss below. This indicates that
the hydrogen column density and dust to gas ratio we used here are
roughly correct without further adjustment.  We note that attenuation
in $r-i$ is a very small quantity, with an amplitude smaller than $0.01$ mag. 
Interestingly, the broad-band photometric data indicates the presence of absorption lines, as shown by the bumps at $z\sim1.2$ and $z\sim1.7$ for the $r-i$ and $i-z$ colours. The small offsets (of order 0.005 magnitude) between observations and the modelled colours might be due to variations in the metal lines as a function of redshift.

We note that if we would take the MW type dust, instead of the
SMC type dust, we cannot reproduce colour excess shown in Figure
3, notably in the $u-i$ for $z>0.6$ and $g-i$ for $z>1.5$. This is
consistent with what has been found in York et al. (2006), and
compel us to adopt the SMC type extinction for \MgII\ absorbers.

In the FUV quasars behind a hydrogen column density with 
log $N_H\gtrsim18$ are usually not detectable by GALEX. The 
colour excess can only be measured for lower column density absorbers. 
We also show in figure ~\ref{fig:color_redshift} 
expected reddening for both absorbers with a column density
somewhat above the Lyman limit: log~$N_H=18.0$, as dashed lines. The overall shape of the extinction shows that this selection effect is well reproduced.

Figure~\ref{fig:reddening_shape} presents the extinction curve derived 
from our \MgII\ sample
(data points) at two redshifts, compared with the one (thick solid curve)
expected from a sum of the dust extinction curve (dashed curve) taken from
\citet{2001ApJ...548..296W} and hydrogen Lyman edge excitation (dotted curve). The curve 
assumes SMC type dust for hydrogen column density $N_{\rm H}=10^{19.8}$ cm$^{-2}$ 
with the dust-to-gas ratio 1/108. For the hydrogen excitation 
the hydrogen column density is
assumed to be  $10^{18}$  cm$^{-2}$, consistent with 
Figure~\ref{fig:color_redshift}, 
avoiding the transmission
being completely opaque. They are close to the case observed 
for our sample, as seen above.
The figure shows that the SMC
extinction curve gives observed extinction with \MgII\ absorbers
correctly for $\lambda>0.12\,\mu$m: extinction at shorter 
wavelengths is properly given by Lyman edge absorption.

\begin{figure}
   \centering
   \includegraphics[width=\hsize]{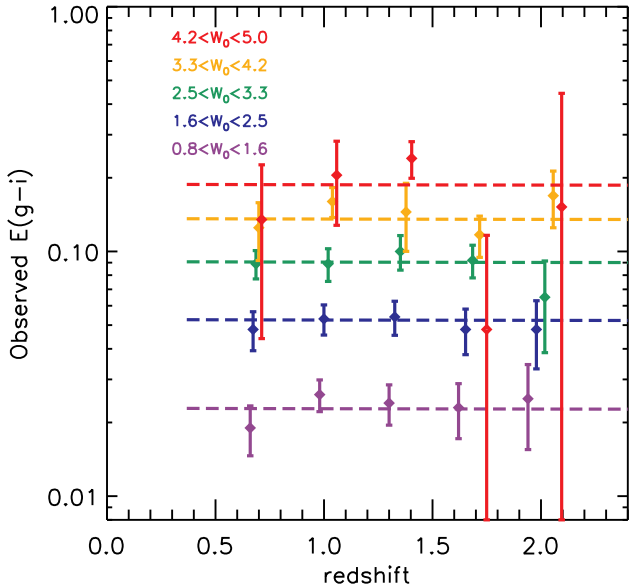}
   \caption{Observed reddening of \MgII\ absorbers as a function of redshift for different bins of rest equivalent widths. From purple to yellow, we select absorber samples with increasing $W_0$ from 0.8 to 5.5 \AA.}
\vspace{.6cm}
   \label{fig:galex}
\end{figure}

Let us focus on dust extinction. The above
figure shows that over the redshift range $0.4<z<2.4$ extinction in the
FUV, NUV and $u$ band is sensitive to  
hydrogen absorption. In order to probe the effect of dust reddening we 
consider the $g-i$ colour. Redder colours are more affected
by metal line absorption. We measure the median colour
excess $E_{g-i}$ as a function of the absorber redshift with the
sample divided into rest equivalent width bins, with 
the results shown in Figure~\ref{fig:galex}. The global variation 
can be fit  with
\begin{equation}
E_{g-i}(W_0,z) = E_{g-i,0}\,\left ( {W_0 \over 1{\rm \AA}}
\right)^\alpha\;(1+z)^\beta,
\label{eq:E}
\end{equation}
where $E_{g-i,0}=0.017\pm0.003$, $\alpha=1.6\pm0.1$, $\beta=-0.01\pm0.22$  and the reduced $\chi^2$ of 0.6. We use this reddening
measurement to infer the cosmic density of dust in \MgII\
absorbers.
Since $N_{\rm HI}\propto W_0^2$ (see equation (\ref{eq:nhi}) below), this
result is consistent, if marginally, with 
$E_{g-i}\propto N_{\rm HI}$, an expected result,
when we take the fact that the $N_{\rm HI}-W_0$ relation has a large scatter
into account.

This may be taken to be a lower limit on 
reddening induced by \MgII\ absorber systems taking into account
the possibility that some highly obscured quasars
might be dropped out of the sample.
Budzynski \& Hewett (2011) estimated from SDSS quasar spectra that 
about 20\% of $W_0>1{\rm \AA}$ \MgII\ absorbers 
are missed due to large obscuration.

\section{The mass density of dust\\ borne by \MgII\ clouds} 
\label{sec:dustdensity}

The comoving density of a population can be written 
\begin{eqnarray}
  n =     {{\rm d} N \over {\rm d} X}\,{1 \over \sigma}\,
\label{eq:density}
\end{eqnarray}
where $\sigma$ is the cross section of the system, 
$d N/d X$ is the number intersected in the interval $X-X+dX$,
and the absorbing distance $X(z)$ is given by
\begin{eqnarray}
{\rm d} X = {c\,(1+z)^2 \over H(z)}\,{\rm d} z\;
\end{eqnarray}
with $H(z)$ being the Hubble constant at redshift $z$.
The comoving number density of the population is then
\begin{eqnarray}
  n &=&   {{\rm d} N \over {\rm d} z}\, 
    { 1 \over dX/dz}\,  {1 \over \sigma}\, .
\label{eq:density}
\end{eqnarray}
A similar relation is derived for the cosmic mass density.
We can write the mass density of dust in \MgII\ absorbers as
\begin{eqnarray}
\rho_{\rm dust}^{\rm MgII}(z) = \left({d N \over d z}\right)_{\rm MgII} \,
{ \Sigma_{\rm dust}(z) \over dX/dz},
\label{eq:rho_dust}
\end{eqnarray}
where $\Sigma_{\rm dust}$ is the surface dust-mass density of \MgII\ absorbers.
In units of the present-day critical density $\rho_{\rm crit}$,
\begin{eqnarray}
\Omega_{\rm dust}^{\rm MgII}(z) = 
{ \rho_{\rm dust}^{\rm MgII}(z)
\over
 \rho_{\rm crit}} .
\label{eq:omega_dust}
\end{eqnarray}
We note that this estimate of $\Omega_{\rm dust}^{\rm MgII}$ does not
require the knowledge of the spatial distribution of \MgII\ absorbers
around galaxies, so that we obtain a lower limit on $\Omega_{\rm
  dust}$ without further knowledge on the distribution of clouds.

The amount of extinction is related to the surface density of dust as,
\begin{eqnarray}
 \Sigma_{\rm dust} =  \frac{{ \ln 10\;A_V}}{2.5\; {K_{{\rm ext},V}} }
\end{eqnarray}
where $K_{{\rm ext},V}$ is the extinction-to-dust-mass coefficient evaluated
at the $V$ band wavelength,
\begin{equation} 
 K_{{\rm ext},V}={{\sigma_{{\rm ext},V}}\over \mu\, m_{\rm H}}
\left(
{\rho_{\rm gas} \over \rho_{\rm dust}}
\right)\,,
\end{equation}
with $\rho$ being the mass density of each species, ${\sigma_{{\rm ext},V}}$ 
the extinction cross section of dust in the $V$ band per hydrogen atom,
 $\mu$ the mean molecular weight of gas, and $m_{\rm H}$ the hydrogen mass. 
The value of $K_{{\rm ext},V}$ is given in a tabulated form for the 
model of \citet{2001ApJ...548..296W} at the electronic address quoted above. 
For SMC type dust we obtain
\begin{equation}
{K_{{\rm ext},V}} 
\simeq 1.54\times10^{4}\,\rm{cm}^2\,\rm{g}^{-1}\,.
\label{eq:K}
\end{equation}
We note that Milky Way type dust leads to a dust mass larger by
a factor of 1.8 at given extinction in the $V$-band.  

\begin{figure}
   \centering
   \includegraphics[width=.9\hsize]{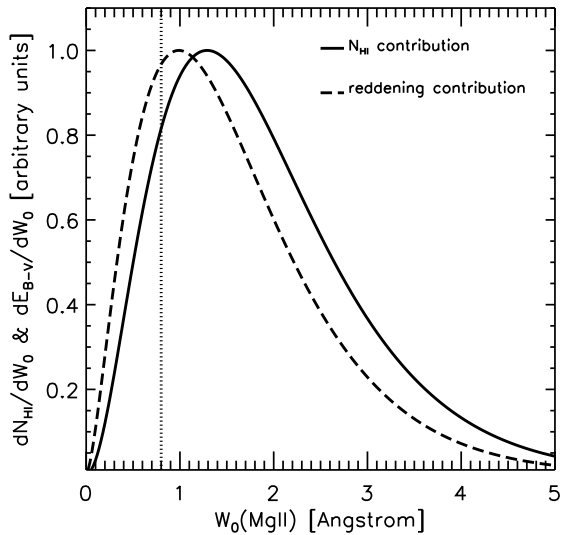}
   \caption{Relative importance of the contribution from
\MgII\  clouds as a function of their rest equivalent widths 
to the integral of dust reddening (dashed line) and total neutral hydrogen
amount  
(solid line). The vertical line is at  $W_0\sim0.8\,{\rm \AA}$, the lower
cutoff of our integral.}
   \label{fig:nhi_contrib}
\end{figure}

The cosmic mass density of dust in \MgII\ absorbers is
\begin{eqnarray}
\rho_{\rm dust}^{\rm MgII}(z) &=& 
{\ln 10 \over {2.5 \, K_{{\rm ext},V}}\,dX/dz}
\int
d W_0 {d^2 N \over d z\,d W_0}\,
 E_{B-V}(W_0,z)\nonumber\\
\label{eq:rho_dust}
\end{eqnarray}
where the incidence of \MgII\ absorbers is given 
by \citet{Nestor+05} from  SDSS quasar spectra,
\begin{eqnarray}
{d^2 N \over d z\,d W_0} = {N^* \over W^*}\,\exp{(W_0/W^*)}
\label{eq:nestor}
\end{eqnarray}
with $N^*= 1.001\pm 0.132\,(1+z)^{(0.226\pm0.170)}$ and
$W^*=0.443\pm0.032\,(1+z)^{0.634\pm0.097}$
\AA.

The integrand of Eq~\ref{eq:rho_dust} is shown by the solid line in
Figure~\ref{fig:nhi_contrib}.  It shows that absorber systems with
$W_0<0.8\,{\rm \AA}$, if included, would contribute by 20\% to
the mass density of dust. The integral above 
$W_0>5\,$\AA\ is about 1\%.
The contribution from obscured quasars may also increase the dust amount
by 20\% if we take literally the effect estimated by 
Budzynski \& Hewett (2011).\\

\begin{figure*}
   \centering
   \includegraphics[width=.7\hsize]{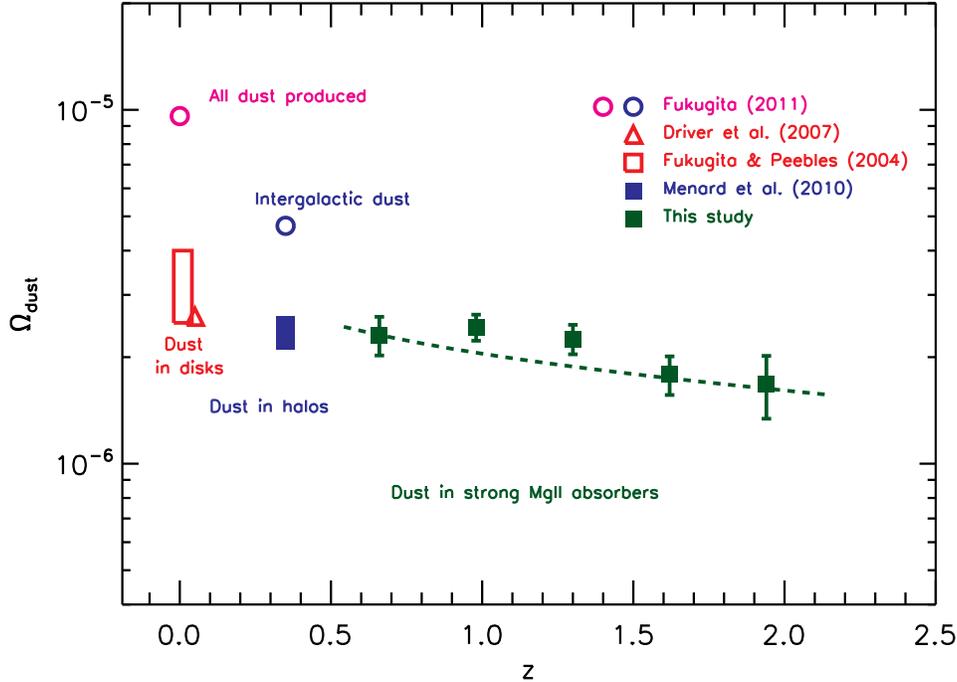}
   \caption{Cosmic mass density of dust contained in \MgII\ absorbers (green
     diamonds) compared with various other estimates of the dust density, 
contained
in galactic disks (Fukugita \& Peebles 2004; Driver et al. 2007), in 
virial radii of galactic haloes (M\'enard et al.), in intergalactic space
beyond virial radii, and the total amount produced in cosmic history
(Fukugita 2011). The dashed curve refers to the estimate from a global
fit using a parametrised form for the redshift dependence for the entire sample 
(see text).}
   \label{fig:omega_dust}
\end{figure*}

Using Equation~\ref{eq:rho_dust} and the 
measured reddening shown in figure~\ref{fig:galex}
we estimate the comic density 
of dust carried by \MgII\ absorbers 
as a function of redshift. The results are shown 
in Figure~\ref{fig:omega_dust}. We find 
$\Omega_{\rm dust}^{\rm MgII} \simeq 1-2\,\times\,10^{-6}$ and an indication
for a slight increase from $z\approx 2$ to 0.5. 
We also estimate $\Omega_{\rm dust}^{\rm MgII}(z)$ using 
Equation~\ref{eq:rho_dust} together with the best fit parameters for Eq.~\ref{eq:E}. The result is shown with the dashed curve. It indicates that
$ \Omega_{\rm dust}^{\rm MgII}\propto (1+z)^\gamma$ 
with $\gamma\simeq{-1}$.

We quote as a representative value at a low redshift ($z\approx 0.5$),
\begin{equation}
\Omega_{\rm dust}^{\rm MgII} \simeq 2.3 \pm 0.2\,\times\,10^{-6}\;.
\end{equation}
As mentioned above, this 
provides us with a lower limit on $\Omega_{\rm dust}^{\rm halo}$ as
our analysis does not include contributions from lines of sight that
(i) produce weak or no \MgII\ absorption, and that (ii) intercept
high column density clouds with the amount of dust that obscures the
background source.
We saw that dust missed for the former reason would increase the amount by
approximately $20\%$.

%
Figure~\ref{fig:omega_dust} includes estimates of the amount of dust
in the literature.
MSFR give the abundance of dust within the virial radius 
of representative galaxies
at $z\approx 0.34$ ($\Omega_{\rm dust}^{\rm halo}\simeq 2.1\times10^{-6}$).
\citet{fukugita2011} estimated the dust mass density within the
virial radius integrating over all galaxies assuming a typical 
luminosity function
and that the dust amount produced is proportional to luminosity, 
$2.6\times10^{-6}$ at 
the median redshift of the sample they used, and 
also extended it to include the component outside the
virial radius, which is indicated by MSFR. The latter gives
$\Omega_{\rm dust}^{\rm IGM}\approx 5\times 10^{-6}$, showing that the total
abundance of dust is consistent with the amount ought to be produced 
in star forming activity in  cosmic time, when added to dust in disks
$\Omega_{\rm dust}^{\rm disk}\approx 3-4\times 10^{-6}$.
It is noteworthy that the dust amount in \MgII\ clouds is close to that 
within the virial radius of galaxies, suggesting that dust is
predominantly held in \MgII\ clouds within the virial radius.
This is approximately 1/2 the total amount 
of dust expected to be spread outside galaxies. 
We note that the calculation of
both $\Omega_{\rm dust}^{\rm halo}$ and $ \Omega_{\rm dust}^{\rm MgII}$
adopts SMC type dust and the same reddening-to-dust mass ratio.

The estimate of Fukugita \& Peebles (2004) refers to dust residing in disks of local galaxies, and Driver et al. (2007) estimate the dust abundance in disc galaxies at $z\approx 0$, both indicated in the figure for comparison. 
This shows that the amount of dust in \MgII\ absorbers is  similar to that remaining in galactic discs: much is blown out of galaxies.

\section{The hydrogen mass density in \MgII\ absorbers}

The same formalism also applies to estimate the mass density of 
neutral hydrogen associated with 
\MgII\ absorbers \citep{1995ApJ...440..435L}. We have
\begin{eqnarray}
\rho_{\rm HI}^{\rm MgII}(z) = 
\left({d N \over d  z}\right)_{\rm MgII} \,
{ N_{\rm HI}\,m_{\rm H} \over dX/dz}\ .
\label{eq:rho_hi}
\end{eqnarray}
The neutral hydrogen column density of \MgII\ absorbers has been studied by 
\citet{2006ApJ...636..610R}, who compiled  about 200 
Lyman-$\alpha$ measurements of \MgII\ absorbers.
Using this sample, \citet{2009MNRAS.393..808M} showed that the median 
$N_{HI}$ of \MgII\ absorbers
is described by
\begin{equation}
N_{\rm HI} (W_0) = (2.45 \pm 0.38) \times 10^{19}\;
\left( {W_0 \over \rm{\rm \AA}} \right)^{2.08 \pm 0.24}\,{\rm cm^{-2}}\;.
\label{eq:nhi}
\end{equation}
Using this relation\footnote{We remark that this gives a column density
of $W_0<2$ \AA~clouds below the empirical threshold of star formation
in galaxies derived by \citet{Kennicutt}.} 
and Equations~\ref{eq:nestor} \& \ref{eq:rho_hi}, we find
\begin{eqnarray}
\Omega_{\rm HI}^{\rm MgII} \simeq (1.5\pm0.3)\times10^{-4}
\end{eqnarray}
for the total sample which has median redshift $z\approx1$
\footnote{
After the completion of this work we became aware
   that \cite{2011ApJ...743L..34K} derived a similar
   (and consistent) estimate of $\Omega_{\rm HI}$ traced by \MgII\ absorbers.
   }. 
The
HI abundance in \MgII\ clouds may slightly decrease towards low redshift,
but the change is within the error.
We thus compute the global dust-to-HI ratio for \MgII\ absorbers,
\begin{eqnarray}
{\Omega_{\rm dust}^{\rm MgII} 
\over
\Omega_{\rm HI}^{\rm MgII} }
\simeq {1 \over 51\pm 15}.
\end{eqnarray}
Including the mass contribution from Helium and heavier elements, it gives a dust-to-gas mass ratio of about $1/(70\pm20)$, which is consistent with the standard value for normal galaxies including the Milky Way.

Let us compare the amount of neutral hydrogen contained in \MgII\ absorbers 
to the cosmic density. The HIPASS HI survey  with 1000 galaxies 
(Zwaan et al. 2003) gives
\begin{eqnarray}
\Omega_{\rm HI}^{\rm HIPASS} \simeq (4.2\pm0.7)\times10^{-4}
\end{eqnarray}
from the  integration of the HI mass function assuming the  Schechter form. 
The objects with $M({\rm HI})>10^7M_\odot$ are all identified with optical
galaxies at least in high latitude fields where one can avoid confusions. 
This suggests that a smooth interpolation of HI objects is also
likely to consist of the small galaxy population, rather than another 
population such as \MgII\ clouds around galaxies,
although the possibility is not excluded that 
it includes contributions from \MgII\ clouds.\footnote{For reference, 
the molecular hydrogen abundance from the CO survey of Keres et al. (2003) is
$\Omega_{\rm H_2}^{\rm tot} \simeq (1.6\pm0.6)\times10^{-4}$.
} Fukugita \& Peebles have estimated that, at the present epoch, the
total cosmic density of HI amounts to
\begin{eqnarray}
\Omega_{\rm HI}^{\rm tot} \simeq 4.5\times10^{-4}\,.
\end{eqnarray}
Our result shows that \MgII\ absorbers add about a third of the 
neutral hydrogen amount in galactic discs and bear a fourth of neutral 
hydrogen amount
in the Universe.

\section{Summary and implications}

Large quasar samples, aided with precision photometry, have enabled us
to study the dust content of \MgII\ absorbers over a broad redshift range.
In addition the combination of high-redshift systems and UV observations allows us to study the short-wavelength extinction properties of their dust grains, all the way to the Lyman edge.

With minimum assumptions we have estimated the
cosmic density of dust borne by \MgII\ clouds. 
At low redshift it amounts to
\begin{equation}
\Omega_{\rm dust}({\rm Mg\,II } ) \approx 2.3\times 10^{-6}\;.
\label{eq:omega_dust_mgii}
\end{equation}
This is comparable to the amount
of dust in galactic disks: $\Omega_{\rm dust}^{\rm disk}\approx
3-4\,\times 10^{-6}$ (Fukugita \& Peebles 2004, Driver et al. 2007).

Fukugita (2011) estimated the total amount of dust produced in the
Universe to be
\begin{eqnarray}
\Omega_{\rm dust}^{\rm tot}&\approx& 0.003~(\Omega_{\rm star}) \times 0.6~ 
({\rm fraction~shed})\nonumber\\
&&\times \, 0.02~({\rm metallicity}) \, \times \,0.3~({\rm fraction~condensed})\nonumber\\
&\approx&
\,1\times 10^{-5}\;.
\end{eqnarray}
This implies that the amount of dust expelled from galaxies is about
$\Omega_{\rm dust}({\rm expelled})\approx 6\,\times 10^{-6}$.
\MgII\ absorbers therefore carry about 1/3 to 1/2 of the total amount of dust expelled by galaxies.

The ratio of the abundances 
$\Omega_{\rm dust}({\rm Mg\,II } ) / \Omega_{\rm dust}({\rm expelled})$
indicates that the gas responsible for \MgII\ absorption 
should have integrated dust for the 
corresponding fraction of the cosmic age:
\MgII\ clouds should have persisted
for a time scale of order several Gyr. The overall gas 
distribution does not disperse too quickly and lasts \emph{effectively}
for a significant fraction of the cosmic age.
The gas distribution traced by \MgII\ absorbers should have received dust produced in nearby galaxies and integrated it at least for a few Gyr period.

In the present study we measured the shape of the extinction curve from
$900\,$\AA, at which the absorption of Lyman edge contributes
predominantly, to longer wavelengths where the opacity is characterized
by dust extinction. The presence of metal absorption lines such as 
\MgII\ and {Fe\,{\sc ii}} is indicated in the broad-band brightness changes due to the presence of absorbers. Larger samples would allow detections of Lyman lines and additional UV metal lines in photometric observations.

We also estimated the hydrogen mass density borne by \MgII\ cloud to
be $\Omega_{\rm HI}\approx 1.5\times 10^{-4}$. This means that the
dust-to-hydrogen mass density is of the order of $1/100$, which
is similar to the fraction for normal galaxies, indicating that
\MgII\ clouds, thought to be devoid of stars and hence have 
no sources of dust, cannot be aggregates of pristine gas,
but are a secondary product from the activity of galaxies. This 
supports their outflow origin, which was
indicated for star forming galaxies \citep{Norman96,Bond01}
and the overall redshift dependence of star formation
\citep{2011MNRAS.417..801M,2012arXiv1201.3919M}. 
Our effective lifetime estimate of the clouds, 
inferred from the abundance consideration, indicates that 
such clouds are not ephemeral but last for a long time,
consistent with the fact 
known from early times that
\MgII\ absorbers are not only associated with star bursting galaxies
but also with galaxies of any morphological types, irrespective
of on-going star-forming activity \citep{SDP94}.
Wherever stars exist, there was bursting phases in the past, and the cloud
can integrate outflows. 

The neutral hydrogen abundance in \MgII\ clouds is 5\% of the mass
density of stars in galaxies, or 10\% of gas shed by stars during
evolution in the cosmic time, assuming the Chabrier initial mass
function (Fukugita 2011). For some actively star-forming galaxy
samples it is observed that a significant fraction of, sometimes as
much as, the star-forming mass is outflowed (e.g., Heckman et al.
2000; Pettini et al. 2002; Veilleux et al. 2005; Rupke et al. 2005;
Weiner et al. 2009). Our estimate of the neutral hydrogen abundance
in \MgII\
absorbers 
also supports these observations of outflows.
The present analysis is an example of the use of dust as a tracer
of mass transactions in the universe with the implication that can be
derived by tracing the fate of dust.

\begin{acknowledgements}
B.M. is supported by the Sloan Foundation, the NSF and the Henri Chr\'etien 
grant.
M.F. thanks Monell Foundation at the Institute for Advanced Study, and receives
Grant-in-Aid of the Ministry of Education in Tokyo.
\end{acknowledgements}


\end{document}